\documentclass[letterpaper,12pt]{article}
\pdfoutput=1
\usepackage[paper=letterpaper,margin=1.0in]{geometry}
\usepackage{amsfonts,amsmath,amssymb}
\usepackage{soul}

\newcommand{\be}{\begin{equation}}
\newcommand{\ee}{\end{equation}}

\newcommand{\bea}{\begin{eqnarray}}
\newcommand{\eea}{\end{eqnarray}}

\newcommand{\nn}{\nonumber}
\newcommand{\Tr}{\text{Tr}}

\newcommand{\rd}{\partial}
\renewcommand{\bar}{\overline}

\def\be{\begin{equation}}
\def\ee{\end{equation}}
\def\bal{\begin{array}{l}}
\def\ba#1{\begin{array}{#1}}  
\def\ea{\end{array}}
\def\bea{\begin{eqnarray}}
\def\eea{\end{eqnarray}}
\def\beas{\begin{eqnarray*}}
\def\eeas{\end{eqnarray*}}

\def\nn{\\\nonumber}

\def\bit{\begin{item}}
\def\eit{\end{item}}
\def\benu{\begin{enumerate}}
\def\eenu{\end{enumerate}}

\makeatletter
\@addtoreset{equation}{section}
\makeatother

\newcommand{\comment}[1]{}

\renewcommand{\thefootnote}{\fnsymbol{footnote}}

\begin{document}

\rightline{WITS-CTP-120}

\vspace*{1.2cm}

\begin{center}

{\LARGE \bf Entanglement Entropy of Extremal BTZ}

\vspace{1cm}

{\large \bf Pawe{\l} Caputa}$^1$\footnote{pawel.caputa@wits.ac.za},
{\large \bf Vishnu Jejjala}$^1$\footnote{vishnu@neo.phys.wits.ac.za},
{\large \bf Hesam Soltanpanahi}$^{1,2}$\footnote{hesam.soltanpanahisarabi@wits.ac.za}

\vspace*{.5cm}

{${}^{1}$ NITheP, School of Physics, and Centre for Theoretical Physics,\\
University of the Witwatersrand, Johannesburg, WITS 2050, South Africa\\
}
\vspace*{.2cm}
{${}^{2}$ Institute of Physics, Jagiellonian University,\\ Reymonta 4, 30-059 Krak\'{o}w, Poland\\
}

\end{center}

\vspace*{1cm}
\centerline{\textbf{Abstract}}
\bigskip

In this note, we use entanglement entropy as a tool to explore the universal properties of CFTs dual to extremal BTZ black holes.
We demonstrate that the entanglement entropies computed in the CFTs at the boundary of the extremal BTZ and the boundary of the near-horizon limit of the extremal BTZ are in perfect agreement and have the form appropriate to a two-dimensional CFT with only the chiral part excited and the anti-chiral in the ground state.
Furthermore, we analyze the universal limits of the entanglement entropy and recover the correct value of the thermal entropy for large entangling intervals and the first-law like relation for the small interval.

\newpage
\renewcommand{\thefootnote}{\arabic{footnote}}
\setcounter{footnote}{0}


\section{\label{sec:intro}Introduction and summary}

Black holes have an entropy, which is non-extensive with the size of the system~\cite{bh}.
Rather, the entropy of a black hole scales with the area of the event horizon.
A theory of statistical mechanics underpins gravitational thermodynamics, and entropy enumerates the quantum configurations of a system that are indistinguishable to an observer who measures global features of the spacetime such as the ADM charges~\cite{sv}.
The holographic principle~\cite{holo}, which contends that the fundamental degrees of freedom of any gravitational system are codimension one, finds a realization in the AdS/CFT duality~\cite{adscft}, for which, by now, there is an overwhelming body of evidence.
Within the strong coupling regime of gravity in asymptotically AdS spaces or for black holes with an AdS factor in the near-horizon region, we may perform detailed calculations using only the perturbative methods of quantum field theory.
Exploiting the correspondence in highly symmetric examples, we can even reconstruct the black hole metric from coarse-graining the microstates in a suitable ensemble~\cite{microstates}.

While we understand that a collection of states in the CFT accounts for black hole entropy, it is not immediately evident how to isolate particular regions in the bulk (for example, the horizon and its associated degrees of freedom) from the dual field theory.
The concept of entanglement entropy provides a language with which to address this question~\cite{Ryu:2006bv}.
(See also~\cite{hms} and~\cite{th} for antecedents.)
Simply stated, the entanglement entropy computes the entropy of a subsystem $A$ upon tracing over its complement $A^c$, which can be regarded as the environment, and is a universal quantity as it is defined in the same manner as the von Neumann entropy.
Entanglement entropy admits a geometric interpretation~\cite{Ryu:2006bv}:
\be
S_A = \frac{\mathrm{Area}[\gamma_A]}{4G_N} ~,\label{RT}
\ee
where the bulk is $(d+2)$-dimensional anti-de Sitter space with Newton constant $G_N$, the CFT lives on $\mathbb{R}^{1,d}$ or $\mathbb{R}\times S^d = A\cup A^c$, and $\gamma_A$ is the $d$-dimensional static minimal surface in AdS$_{d+2}$ whose boundary is $\partial A\subset \mathbb{R}^d$ or $S^d$. 
A host of calculations --- see~\cite{Nishioka:2009un} for a review --- buttresses the relation between entanglement entropy and black hole entropy. 
For sufficiently large interval $A$, the minimal surface in the Ryu--Takayanagi formula encircles the black hole horizon and the dominant contribution to \eqref{RT} reproduces the Bekenstein--Hawking entropy.

In this article, we investigate the entanglement entropy of extremal BTZ black holes using, firstly, the CFT dual associated to the AdS$_3$ which the black hole inhabits and, secondly, the CFT dual associated to the AdS$_3$ factor in the near-horizon region.
While the metric for an extremal BTZ black hole may be obtained as the limit of the non-extremal BTZ black hole wherein the inner and outer horizons are equated, mapping the physics of one onto the other is an extremely subtle business.

The near-horizon limit of an extremal black hole leads to a spacetime metric that solves the same Einstein equations as the original solution~\cite{reall}.
Moreover, it may have enhanced symmetries.
In the case of the extremal Kerr solution, for example, the isometry group is enhanced from $\mathbb{R}\times U(1)$ to $SL(2,\mathbb{R})\times U(1)$~\cite{barhor}.
The Kerr/CFT correspondence posits that in the near-horizon limit of the extremal Kerr black hole in four dimensions, the Cardy formula reproduces the microscopic entropy~\cite{Guica:2008mu}.\footnote{
See~\cite{vjsn} for a discussion of some of the subtleties that accompany the use of the Cardy formula.}
The degrees of freedom in the near-horizon limit are identified with a chiral half of a two-dimensional CFT with central charge determined by the angular momentum $J$.
The same applies to the extremal BTZ black hole:
the near-horizon limit is dual to the DLCQ of a $(1+1)$-dimensional CFT and manifests a chiral Virasoro algebra~\cite{Balasubramanian:2009bg}.
The authors of~\cite{Azeyanagi:2007bj} argue that the entropy of extremal and near-extremal black holes can be interpreted in terms of the entanglement entropy of a $(0+1)$-dimensional CFT (\textit{viz.}, conformal quantum mechanics).
The recent identification of a first law-like relation for the entanglement entropy allows us to calculate the dimension in which the CFT resides.

Let us state our results at the outset.
A careful computation of the entanglement entropy in the CFT$_2$ associated to the asymptotic region of the extremal BTZ spacetime exactly matches the entanglement entropy in the CFT$_2$ associated to the boundary of the near-horizon AdS$_3$.

These calculations are couched purely in terms of CFT. It is well known that taking the extremal and near horizon limits is a subtle procedure (see \textit{e.g},~\cite{Balasubramanian:2009bg}). Therefore, we have carefully performed two computations in the two above mentioned limits. Firstly, we closely follow~\cite{Hubeny:2007xt} that derived entanglement entropy in left/right asymmetric ensembles and matched this to the gravity computation in the non-extremal spinning BTZ solution. We take the extremal limit of the BTZ metric and compute the entanglement entropy in the CFT dual. Next, we carefully take the near horizon limit of the extremal solution and compute the entanglement entropy in the corresponding CFT dual as well.  The second calculation should be viewed  as complementary to gravity results in warped AdS$_3$ spaces based on the covariant Ryu--Takayanagi prescription performed recently in~\cite{Anninos:2013nja}. We find a perfect agreement between two answers. 

We would like to emphasize that the former CFT lives at the asymptotic boundary (UV limit) while the later one lives in the near horizon region (IR limit). \textit{A priori}, there is no reason to find the same results from these two different points of view.
Using the first law-like relation that correlates the change in energy to the change in the entanglement entropy, we conclude the CFT is $(1+1)$-dimensional.
Thus, we interpret the degrees of freedom as corresponding to one chiral half of a two-dimensional CFT.

The outline of this paper is as follows.
In Section~\ref{sec:2}, we define the ultraviolet CFT as the field theory that lives on the boundary of the AdS$_3$ in which we find the BTZ black hole and the infrared CFT as the field theory that lives on the boundary of the AdS$_3$ factor in the near-horizon limit.
In Section~\ref{sec:3}, we compute the entanglement entropy in the ultraviolet and infrared CFTs using the replica trick for an interval of length $L$ on the boundary.
In Section~\ref{sec:4}, we take the large $L$ and small $L$ limits of the formula for the entanglement entropy.
The large $L$ limit recovers the Frolov--Thorne temperature and the Bekenstein--Hawking entropy of the extremal BTZ solution as its leading term.
The small $L$ limit verifies that the CFT is two-dimensional, and the excitations that we are examining reside in one chiral sector of the theory.
In Section~\ref{sec:5}, we compare our purely CFT calculations to holographic computations in the literature.
Section~\ref{sec:6} then offers a discussion and a prospectus for our future work in this direction.
For completeness, Appendix~\ref{app:A} reviews the calculation of the entanglement entropy in the field theory dual of the non-extremal spinning BTZ black hole.
Appendix~\ref{app:B} computes the energy-momentum tensor associated to an extremal BTZ spacetime in the near-horizon geometry.

\section{\label{sec:2}Extremal BTZ} 

To fix notation, we briefly recapitulate the near-horizon limit of the extremal BTZ black hole as well as the map from this geometry to AdS$_3$.
Recall that the metric of the rotating BTZ black hole~\cite{Banados:1992wn} is
\be
\mathrm{d}s^2 =-\frac{(r^2-r^2_+)(r^2-r^2_-)}{R^2r^2}dt^2+\frac{R^2r^2\,dr^2}{(r^2-r^2_+)(r^2-r^2_-)}+r^2(d\phi-\frac{r_+r_-}{Rr^2}dt)^2 ~,
\label{neBTZ}
\ee
where $R$ is the radius of anti-de Sitter space and $r_{\pm}$ are the radii of the outer and the inner horizons.
The coordinate $\phi$ is an angular variable with period $2\pi$.
The mass $M$, the angular momentum $J$, and the Hawking temperature $T_H$ of the black hole are expressed in terms of the radii as
\be
MR^2=r^2_++r^2_- ~, \qquad
JR=2r_+r_- ~, \qquad
T_{H}=\frac{r^2_+-r^2_-}{2\pi R^2 r_+} ~.
\ee
We realize the extremal limit in which $M\, R=J$ by putting $r_0\equiv r_+=r_-$ in~(\ref{neBTZ}).
In this limit, the Hawking temperature vanishes.
The metric then reduces to the expression
\be
\mathrm{d}s^2=-\frac{(r^2-r^2_0)^2}{R^2r^2}dt^2+\frac{R^2r^2}{(r^2-r^2_0)^2}dr^2+r^2(d\phi-\frac{r^2_0}{Rr^2}dt)^2 ~.
\label{extremal-metric}
\ee
This solution can be mapped to Poincar\'e form
\be
\mathrm{d}s^2=R^2\frac{dU^2+d\omega_+ d\omega_-}{U^2}
\label{AdS3}
\ee
by the coordinate transformation~\cite{KeskiVakkuri:1998nw}
\be
w_- = \frac{R}{2r_0}e^{\frac{2r_0}{R}(\phi-t/R)} ~, \qquad
w_+ = \phi+\frac{t}{R}-\frac{Rr_0}{r^2-r^2_0} ~, \qquad
U = \frac{R}{\sqrt{r^2-r^2_0}}e^{\frac{r_0}{R}(\phi-t/R)} ~. \label{mapE} 
\ee
We will show that this map contains sufficient information to compute entanglement entropy in the dual conformal field theory that lives on the asymptotic boundary of~\eqref{extremal-metric}.
We refer to this class of theories as \textit{ultraviolet} conformal field theories.

\subsection{Near-horizon limit of extremal BTZ} 

The near-horizon geometry of the extremal BTZ solution has a conformal field theory description via the standard AdS/CFT dictionary.
As the near-horizon region comprises a warped AdS$_3$, this is also a two-dimensional quantum field theory.
The exact map between the CFT from the near-horizon regime to the CFT that lives on the asymptotic boundary of the spacetime is not known.
We will refer to the former class of theories associated to the near-horizon region as \textit{infrared} conformal field theories.

Let us now review the scaling limit of the extremal metric~\eqref{extremal-metric} and its map to AdS$_3$.
We shall later make use of this to evaluate entanglement entropy in infrared CFTs.

For our purposes, it is convenient to use the following set of coordinates \cite{Carlip:2005zn}:
\be
u_{\pm}=\phi\pm t/R ~, \qquad
r^2-r_0^2=R^2\,e^{\frac{2x}{R}} ~,
\label{coord1}
\ee
in which the metric becomes
\be
\mathrm{d}s^2={r_0^2}\,du_-^2+d x^2+R^2\,e^{\frac{2x}{R}}\,du_+ du_- ~, \qquad
u_{\pm}\sim u_\pm+2\pi ~.
\ee
Both the variables $u_{\pm}$ are periodic.
On the boundary at $x\rightarrow \infty$, $du_{\pm}$ are null directions.
The asymptotic symmetry of this metric is associated with the arbitrary reparametrization invariance by which $u_{\pm}\mapsto f_{\pm}(u_{\pm})$.
Form the point of view of the ultraviolet CFT that lives at the asymptotic boundary, this symmetry corresponds to two chiral Virasoro algebras.

In the new set of coordinates~\eqref{coord1}, the near-horizon limit is defined as follows.
We first introduce the coordinates $y,v_+,v_-$ specified in terms of $x,u_+,u_-$ and a scaling parameter $x_0$:
\be
y=x-x_0 ~, \qquad
v_+=\frac{R\,e^{\frac{2x_0}{R}}}{r_0}\,u_+ ~, \qquad
v_-=\frac{r_0}{R}\,u_- ~.
\ee
The periodicities of the $v_\pm$ coordinates are given by
\be
\{v_+,v_-\}\sim\{v_++2\pi R\frac{e^{\frac{2x_0}{R}}}{r_0}, v_-\} \sim \{v_+,v_-+2\pi \frac{r_0}{R}  \} ~.
\ee
Next, we take the limit $x_0\rightarrow-\infty$ while keeping $y, v_{\pm}$, and $r_0$ fixed.
The near-horizon geometry of extremal BTZ is given by
\be
\mathrm{d}s^2=R^2\,dv_-^2+dy^2+R^2\,e^{\frac{2y}{R}}  dv_+  dv_- ~.
\label{NH1}
\ee
Notice, that in this limit, $v_+$ is decompactified while $v_-$ remains periodic:
\be
v_-\sim v_-+2\pi \frac{r_0}{R}
\ee
The boundary at $y\rightarrow \infty$ is a null cylinder.
Importantly, the metric on the boundary is conformal to $dv_+ dv_-$.
In the dual CFT, one periodic  direction $v_-$ is associated to a single chiral Virasoro algebra based on this coordinate, $\rd_{v_-}\sim \bar{L}_0$.\footnote{
This is in agreement with the investigation of the putative CFT dual to the near-horizon extremal Kerr geometry~\cite{Guica:2008mu}.}

Finally, the metric is written in a familiar warped AdS$_3$ form (\textit{i.e.}, as AdS$_2\times S^1$) by changing to another radial coordinate $\rho=e^{\frac{2y}{R}}$:
\be
\mathrm{d}s^2=\frac{R^2}{4}\,\left(-\rho^2\,dv_+^2+\frac{d\rho^2}{\rho^2}\right)+{R^2}\,\left(dv_- +\frac{\rho}{2}\, dv_+\right)^2 ~, \qquad
v_-\sim v_-+2\pi \frac{r_0}{R} ~. \label{NH2}
\ee
In these coordinates, $\rho$ and $v_\pm$ are dimensionless while the parameters $R$ and $r_0$ have dimensions of length.

The entropy as computed by the full metric~\eqref{extremal-metric} and the entropy in its near-horizon limit~\eqref{NH2} are the same.
That is to say, the Bekenstein--Hawking entropies, proportional to the the area of the horizon, $\text{Area}=2\pi r_0$, is identical for the two spacetimes.
Equivalently, the isometry group of the near-horizon metric is $SL(2,\mathbb{R})\times U(1)$.
Consistent with the Kerr/CFT approach~\cite{Guica:2008mu}, the microscopic entropy obtained from the Cardy formula is expressed in terms of the Frolov--Thorne temperature
\be
S=\frac{\text{Area}}{4G_N}=\frac{\pi^2}{3}\, c \,T_\mathrm{FT} ~,\label{SFT}
\ee
where the central charge is determined by the Brown--Henneaux result $c=\frac{3R}{2G_N}$ and Frolov--Thorne temperature~\cite{Frolov-Thorne} is  $T_\mathrm{FT}=\frac{r_0}{\pi R}$.

Finally, the coordinate transformation
\be
\omega_-=\frac{1}{2} e^{2 v_{-}} ~, \qquad
\omega_+=v_+-\frac{1}{\rho} ~, \qquad
U=\frac{e^{v_{-}}}{\sqrt{\rho}} \label{mapNH}
\ee

brings the near-horizon metric of the extremal BTZ black hole to the Poincar\'e patch of $AdS_3$. Notice that while these coordinate transformations are very similar to the coordinate transformation~\eqref{mapE} for the entire metric, crucially the asymptotic regions are different.
We will make use of this map in our computation of the entanglement entropy in the infrared CFT dual to the near-horizon limit of the extremal BTZ solution.

\section{\label{sec:3}Computation of the entanglement entropy}
In this section we briefly review the computation of the entanglement entropy in a two-dimensional CFT (we refer to~\cite{EE} for a more complete review).
We then compute entanglement entropy in a class of conformal field theories dual to the extremal BTZ black hole (\textit{i.e.}, in the ultraviolet CFT) as well as in the CFT dual to the near-horizon limit of the extremal BTZ solution (\textit{i.e.}, in the infrared CFT).

\subsection{Entanglement entropy}

Entanglement entropy is a function of a particular quantum state.
It measures how a subset of a system in this given state is entangled with its complement.
More pragmatically, given the density matrix $\rho$ associated to a state, we may divide it into two parts, $A$ and its complement $A^c$, and obtain a reduced density matrix $\rho_A$ by tracing over $A^c$.
The entanglement entropy of the region $A$ is the von Neumann entropy of the reduced density matrix:
\be
S_\textrm{EE}=-\Tr\, \rho_A\log\rho_A ~.
\ee
In $(1+1)$-dimensional conformal field theories, the entanglement entropy is neatly computed using the replica trick~\cite{Holzhey:1994we,Calabrese:2004eu,Calabrese:2009qy}.
We first define the Renyi entropy $S^{(n)}_A$ as
\be
S^{(n)}_A=\frac{1}{1-n}\log\Tr\left(\rho^n_A\right) ~,
\ee
and then analytically continue to non-integer values of $n$.
The entanglement entropy is given by the limit
\be
S_\mathrm{EE}=-\lim_{n\to 1}\partial_n \Tr(\rho^n_A)=\lim_{n\to 1}S^{(n)}_A ~.
\ee
As shown in~\cite{Calabrese:2004eu}, in the path integral formalism, evaluating $\Tr(\rho^n_A)$ is equivalent to computing the partition function $Z_n(A)/Z^n$, on the $n$-sheeted Riemann surface with cuts corresponding to the subregion $A$.
This partition function is proportional\footnote{
In this work, we ignore the proportionality constant.
This does not affect any of the results.}
to the correlation function on the plane of the twist (anti-twist) fields $\phi^+$ $(\phi^-)$ inserted at the boundary points of the interval $A=[u,v]$.
By this chain of reasoning, the entanglement entropy is determined as
\be 
S_\mathrm{EE}=-\lim_{n\to 1}\partial_n \langle\phi^{+}(u)\phi^{-}(v) \rangle ~,
\ee
where the conformal dimension of the twist fields is equal to
\be
h=\bar{h}=\frac{c}{24}\left(n-\frac{1}{n}\right) ~.
\ee

For example, by following the procedure outlined above and using the standard form of the two-point function constrained by the conformal symmetry, one straightforwardly derives the entanglement entropy of a single interval $L$ in a CFT on the plane~\cite{Calabrese:2004eu}
\be
S_\mathrm{EE}=\frac{c+\bar{c}}{6}\log\frac{L}{\epsilon} ~,
\ee
where $c$ and $\bar{c}$ are the central charges of the left and right sectors, respectively, and $\epsilon$ is the lattice cutoff.
This is the leading divergence, but it receives corrections~\cite{EE}.\footnote{
If we regard the entanglement entropy as counting degrees of freedom, the Cardy formula~\cite{cardy} for microscopic entropy also receives corrections~\cite{shahin}.}
The ultraviolet cutoff $\epsilon$ can be thought of as the regulator of the twist fields.

\subsection{Entanglement entropy in a CFT dual to extremal BTZ}

Our goal is to calculate the entanglement entropy of a single interval in a CFT at the asymptotic boundary of the extremal BTZ background and compare to a similar calculation in the CFT at the asymptotic boundary of the near-horizon limit.
As we show below, even without knowing the precise details of these theories, we can nevertheless compute the entanglement entropy using standard holography.
The entanglement entropy within the ultraviolet CFT matches the entanglement entropy within the infrared CFT.

Recall that the CFT dual to black holes generally exist at finite temperature.
They therefore correspond to some general cylindrical geometry at the boundary.
In order to compute the entanglement entropy in such theories we only need the correlator of twist fields.
This can be obtained by inserting an appropriate plane-to-cylinder-like map $w(z)$ to the general transformation of the correlation functions in the CFT:
\be
\langle {\cal O}(z_1,\bar{z}_1) {\cal O}(z_2,\bar{z}_2)\ldots \rangle=
\prod_i \left(\frac{dw_i}{dz_i}\right)^h \left(\frac{d\bar w_i}{d\bar z_i}\right)^{\bar h} \langle  {\cal O}(w_1,\bar{w}_1) {\cal O}(w_2,\bar{w}_2) \ldots \rangle ~.
\label{corr-trans}
\ee
We then follow the usual steps in the replica trick to obtain the entanglement entropy.
The key to this result is isolating the appropriate map $w(z)$.

Given a general black hole, it is not obvious what the map from the plane to a complicated (twisted) cylinder on which the dual theory lives.
Fortunately, as we saw in the previous section, the metrics~\eqref{extremal-metric} and~\eqref{NH2} can be mapped to the Poincar\'e patch of AdS$_3$ by~\eqref{mapE} and~\eqref{mapNH}, respectively.
By taking the asymptotic limit of these maps,\footnote{These are the limits $r\to \infty$ in~\eqref{mapE} and $\rho\to\infty$ in~\eqref{mapNH}.} we can read off the conformal map between the plane on which the CFT dual to AdS$_3$ lives and the general twisted cylinders (see \cite{KeskiVakkuri:1998nw} for similar arguments in the context of two-point functions). 

Let us first apply this strategy to the map~\eqref{mapE}.
Taking $r\to \infty$ we can read off the transformation from the conformal field theory on the plane to the ultraviolet conformal field theory dual to extremal BTZ.
More precisely, after Wick rotating $t\to i\tau$, we obtain the transformation 
\bea
w(z)&\equiv&\phi+i\frac{\tau}{R}= z ~, \nonumber \\
\bar{w}(\bar{z})&\equiv& \frac{R}{2r_0}e^{\frac{2r_0}{R}(\phi-i\tau/R)}= \frac{R}{2r_0}e^{\frac{2r_0}{R}\bar{z}}\label{Trans} ~.
\eea
The left moving modes remain unchanged whereas the right moving modes transform in a non-trivial manner.
Inserting into~\eqref{corr-trans}, we then derive the two-point function of the twist fields in the ultraviolet CFT:
\bea
\langle \phi^+(z_2,\bar{z}_2)\phi^-(z_1,\bar{z}_1)\rangle
&=&\left(\frac{z_2-z_1}{\epsilon}\right)^{-2h}\left(\frac{R}{r_0\epsilon}\sinh\left(\frac{r_0}{R}(\bar{z}_2-\bar{z}_1)\right)\right)^{-2\bar{h}} ~.
\eea
For the entangling interval of length $L$, we set the boundary points to
\be
z_1=\bar{z}_1=0 ~,\qquad
z_2=\bar{z}_2=L ~.
\ee
The entanglement entropy becomes
\be
S_\mathrm{EE}=\frac{c}{6}\log\left(\frac{L}{\epsilon}\right)+\frac{\bar{c}}{6}\log\left(\frac{R}{r_0\epsilon}\sinh\left(\frac{r_0\,L}{R}\right)\right) ~. \label{eBTZEE}
\ee

Equation~\eqref{eBTZEE} is one of our main results.
This form of the entanglement entropy has a clear interpretation.
The left movers are in their ground state.
The entanglement entropy for the right movers has the thermal form\footnote{
Entanglement entropy in a $(1+1)$-dimensional CFT at finite temperature $T=\beta^{-1}$ is given by $$S_\mathrm{EE}=\frac{c}{3}\log\left(\frac{\beta}{\pi\epsilon}\sinh\frac{\pi L}{\beta}\right) ~.$$
See Appendix~\ref{app:A} for the derivation.} with effective temperature 
\be
T_\mathrm{eff}=\frac{r_0}{\pi R} ~, \label{eq:Teff}
\ee
which recovers the Frolov--Thorne temperature reviewed in the previous section.
As we will see in the next section, this equivalence is in agreement with the universal properties of the entanglement entropy and its relation to the thermal entropy of a CFT~\cite{Calabrese:2009qy}.
It is crucial to note that $T_\mathrm{eff}$ in~\eqref{eq:Teff} is not a real temperature ---
as we are discussing an extremal black hole, the Hawking temperature necessarily vanishes.
The effective temperature measures the excitation in a particular sector of the CFT.
It can be thought of as a Lagrange multiplier that fixes the energy.
This is similar to what happens when counting half-BPS operators with conformal dimension of order $N^2$ in ${\cal N}=4$ super-Yang--Mills theory~\cite{babel}.

Now, the same strategy may be applied to the near-horizon limit of the extremal BTZ black hole.
The difference is that metric~\eqref{NH2} is locally AdS$_3$ in the asymptotic region whereas the complete extremal BTZ is asymptotically AdS$_3$.
Nevertheless, the idea we advanced earlier applies.
The only ingredient that we need in order to compute the entanglement entropy in a class of dual CFTs is the asymptotic map between the plane and a twisted cylinder.
It is evident from~\eqref{mapNH} that taking the limit of $\rho\to\infty$ and redefining $(z,\bar{z})=(v_+,v_-)$ precisely matches the map~\eqref{Trans}.
As a result, the entanglement entropy computed in the infrared CFT dual to the near-horizon limit of the extremal BTZ is the same as~\eqref{eBTZEE} in the ultraviolet CFT.
We will comment further on this result in the light of the RG flow between the two CFTs in Section~\ref{sec:6}.

\section{\label{sec:4}Universal limits}

In this section we further analyze the universal behavior of the entanglement entropy derived above. 
In the large $L$ limit we reproduce the Bekenstein--Hawking entropy of the extremal BTZ black hole, and in the small $L$ limit we derive the first law-like relation from both the extremal BTZ metric as well as the warped $AdS_3$.
The first law-like relation provides further confirmation that the dual theories are chiral parts of a two-dimensional CFT. 

\subsection{Large $L$ limit}
Once the entangling region becomes large, the entanglement entropy becomes extensive and we recover the thermal entropy of the system (see~\cite{EE} for further details).
More precisely, in the limit of $L\to \infty$ the entanglement entropy becomes
\be
S_\mathrm{EE}\to S_\mathrm{th}\, L +C(\epsilon) ~,
\ee
where $S_\mathrm{th}$ is the thermal entropy (equal to the Bekenstein--Hawking entropy of the horizon in the dual geometry in holographic CFTs), and $C(\epsilon)$ is some divergent constant.

Expanding the result~\eqref{eBTZEE} for large $L$ yields 
\bea
S_\mathrm{EE}&\simeq& \frac{c}{6}\log\left(\frac{L}{\epsilon}\right)+\frac{\bar{c}}{6}\log\left(\frac{R}{r_0\epsilon}\frac{1}{2}e^{\frac{r_0}{R}L}\right)\nn\\
&=&\frac{c}{6}\log\left(\frac{L}{\epsilon}\right)+\frac{\bar{c}}{6}\log\left(\frac{R}{2r_0\epsilon}\right)+\frac{\bar{c}\,r_0L}{6R} ~.
\eea
We can read off the value of the thermal entropy
\be
S_\mathrm{th}=\frac{\pi\bar{c}r_0}{3R} ~.
\ee
This is a function of the effective temperature (equal to $T_\mathrm{FT}$) and matches~\eqref{SFT}, the Bekenstein--Hawking entropy of the extremal BTZ black hole.\footnote{We recall that a factor of $2\pi$ has to be inserted to the distance $2\pi L$ when we compare with the thermal entropy of the CFT of the circle at the boundary of the BTZ~\cite{Takayanagi:2012kg}.} 
Note also that for our discussion of the universal properties of the entanglement entropy, we assume the size of the thermal circle to be large. For a finite circle one should, in principle, carefully analyze the Lieb--Araki inequality and the homology constraint.
See the discussion in~\cite{Hubeny:2013gta}), but this is beyond the scope of this paper.

\subsection{Small $L$ limit and the first law}
In the limit of a small entangling region $L$, entanglement entropy obeys the analog of the first law of thermodynamics~\cite{Bhattacharya:2012mi} (see also,~\cite{Blanco:2013joa,Wong:2013gua,Nozaki:2013vta,Allahbakhshi:2013rda,Caputa:2013eka}).
Thus, we have a statement that
\be
\Delta E=T_\mathrm{ent}\,\Delta S_\mathrm{EE} ~,\label{Linear}
\ee
where the change in energy in the interval $L$ is obtained by integrating the holographic energy-momentum tensor $T_{tt}$ over the entangling interval
\be
\Delta E=\int_{L}dx\ T_{tt} ~,
\ee
and the change in entanglement entropy $\Delta S_\mathrm{EE}$ is a difference between the entanglement entropy computed in an excited state and in the vacuum.
The constant of proportionality $T_\mathrm{ent}$ is universal and depends on the number of dimensions of the CFT as well as the shape of entangling interval.
For a $d$-dimensional sphere, $T_\mathrm{ent}$ has the following form:
\be
T_\mathrm{ent}=\frac{d+1}{\pi L} ~.\label{Tent}
\ee
Expanding our result for small $L$, we obtain the expression
\be
S_\mathrm{EE}\simeq \frac{c+\bar{c}}{6}\log\left(\frac{L}{\epsilon}\right)+\frac{\bar{c}\,r^2_0L^2}{36}+O(L^4) ~.
\ee
The difference between the ground state (AdS$_3$) entropy is then
\be
\Delta S=\frac{\bar{c}\,r^2_0L^2}{36} ~.
\ee

From the Fefferman--Graham expansion, we see that the time-time component of the energy-momentum tensor is
\be
T_{00}=\frac{2r^2_0}{16\pi G_N} ~.
\ee
We then calculate that the energy of the entangling interval is
\be
\Delta E=\int_L dx\ T_{00} =\frac{2r^2_0 L}{16\pi G_N} ~.
\ee
Using the Brown--Henneaux relation $R/G_N=2\bar{c}/3$, we finally obtain
\be
\Delta E=\frac{2r^2_0 L}{16\pi G_N}=\frac{3}{\pi L}\Delta S ~.
\ee

The first law-like relation holds for in this CFT is which only the left movers are excited.
Assuming the validity of~\eqref{Tent} that was derived without any crucial assumptions about the excited state of the field theory, it is clear that the CFT under consideration lives in $d=2$.
Note, however, that we only have excitations in one sector as the black hole is extremal. In other words the contribution to the linear relation comes only from the exited left-movers. Functionally, one chiral half of a two-dimensional CFT is dynamically equivalent to conformal quantum mechanics~\cite{sen}.

It is also interesting to derive the first law relation from the warped AdS$_3$ metric.
A puzzle that one faces is that in the holographic stress tensor of the near-horizon metric, there is no $T_{tt}$ component.
Nevertheless it is easy to see that the information about the mass of the extremal BTZ black hole is encoded in the $T_{\varphi\varphi}$ component (see Appendix~\ref{Tvv}).
Hence, we obtain $\Delta E$ by integrating $T_{\varphi\varphi}$ over the interval $L$.

\section{\label{sec:5}Holographic computation with covariant prescription}
Entanglement entropy in conformal field theory can be computed using an AdS/CFT prescription first introduced by Ryu and Takayanagi~\cite{Ryu:2006bv}.
For a single interval $L = [a,b]$ in a $(1+1)$-dimensional CFT, the entanglement entropy is proportional to the length of the minimal geodesic $\mathcal{L}(\gamma)$ with endpoints at $a$ and $b$:
\be
S_\mathrm{EE}=\frac{\mathcal{L}(\gamma)}{4G_N} ~.
\ee
Recently a formal proof of this formula was given by \cite{Lewkowycz:2013nqa} (see also \cite{Caputa:2013eka} for an intuitive derivation in AdS$_3$/CFT$_2$). In the more general case where the entangling interval does not lie in a single time slice of the boundary, the Ryu--Takayanagi formula is generalized to a covariant prescription in~\cite{Hubeny:2007xt}.
In these cases the minimal length $\mathcal{L}(\gamma)$ is replaced by the length of an extremal geodesic.

The covariant Ryu--Takayanagi prescription has recently been applied in the context of warped AdS$_3$ backgrounds in~\cite{Anninos:2013nja}.
One of the examples of the warped AdS$_3$ geometries is the near-horizon limit of the extremal BTZ solution~\eqref{NH2}.
The authors computed the entanglement entropy of a single interval $L$ in this background.
This should correspond to the entanglement entropy of the infrared CFT that we have discussed above.
Their result precisely matches with our formulas obtained for both the ultraviolet and infrared CFTs~\eqref{eBTZEE} and confirms the applicability of the covariant prescription of~\cite{Hubeny:2007xt} in this case.
Moreover,~\cite{Hubeny:2007xt} holographically computes the entanglement entropy in ultraviolet CFTs dual to the general rotating BTZ.
The result in~\cite{Hubeny:2007xt} also agrees with the near-horizon result of~\cite{Anninos:2013nja} as well as the two computations in our paper.

\section{\label{sec:6}Discussion and outlook}
As we demonstrated, our two computations of entanglement entropy in the ultraviolet and infrared CFTs dual to extremal BTZ and its near-horizon limit, respectively, are in perfect accord and also match the holographic results of~\cite{Hubeny:2007xt} and~\cite{Anninos:2013nja}.
According to the standard holographic dictionary, one may expect that the two CFTs under consideration are connected by an RG flow~\cite{mn}.
Since under renormalization we integrate out degrees of freedom as we flow from the ultraviolet to the infrared, it is surprising that the entanglement entropy is equal in the two regimes.
Intuitively, one would expect that after each RG step, a measure of information like entanglement entropy will decrease.
As discussed in~\cite{Swingle:2009bg} this is not true in general and there is still room for understanding the mechanisms behind the monotonicity of entanglement entropy under holographic renormalization.
We hope that progress in this direction can be achieved by understanding the structure of entanglement entropy in CFTs dual to extremal black holes and their near-horizon limits and we return to this issue in future work~\cite{WIP} (see also~\cite{Liu:2012eea} for a related discussion).

Interpreting entanglement entropy as black hole entropy, we note that in a two-dimensional CFT, the entropy receives contributions from both the left and right moving sectors.
For an extremal black hole, the entropy arises as a consequence only of one sector.
This sector is accessed by the near-horizon geometry and the DLCQ limit of the two-dimensional field theory~\cite{Balasubramanian:2009bg}.\footnote{
We thank Shahin Sheikh-Jabbari for a discussion on this point.}

The linear relation~\eqref{Linear} expresses the difference in entanglement entropy computed for the same entangling region but in two different states (in general, both are excited) in terms of the energy associated with the interval.
As we showed, the universality of the proportionality coefficient $T_\mathrm{ent}$ allows us to pin down the dimension of the dual chiral part of a CFT in $d=2$.
This is to be compared with~\cite{Azeyanagi:2007bj}, who examine the extremal BTZ solution in terms of the AdS$_2$/CFT$_1$ correspondence.
Again, because of the similarity between conformal quantum mechanics and one chiral half of a two dimensional CFT, the dynamics are equivalent.

The first law-like relation allows us to extract much deeper information about the connection between entanglement and holography.
Namely, the energy of the entangling interval is computed as the integral over the appropriate component of the stress tensor that is directly linked to the gravity dual via the Fefferman--Graham expansion.
As shown in~\cite{Nozaki:2013vta,Bhattacharya:2013bna}, via the linear relation, one can translate Einstein's equations in the bulk to differential equations that govern the dynamics of the entanglement entropy ($\Delta S$) in the dual CFT.
It would be very interesting to study these equations in detail in the context of extremal BTZ black holes and warped AdS$_3$ spaces that appear as their near-horizon limits and scan for possible patterns that appear in the Kerr/CFT correspondence. 

The $M=0$ BTZ black hole is the effective geometry associated to the D$1$-D$5$ system~\cite{bksadm}.
The boundary CFT here is an ${\cal N}=(4,4)$ supersymmetric sigma model with target space $(T^4)^N/S_N$, where $N$ is the product of the number of D$1$-branes and D$5$-branes~\cite{orb}.
At the orbifold point, we have some control within the CFT and compute correlation functions using twist operators.
Although the horizon classically vanishes, this geometry supplies a useful testbed for future investigations.
In particular, it is in principle possible to compute minimal surfaces in microstate geometries with AdS$_3$ asymptopia.
This is also work in progress~\cite{WIP}.

Finally, there are two obvious directions in which we want to pursue our studies of entanglement entropy in the context of extremal black holes.
First, we have the higher spin black holes (see the review~\cite{Ammon:2012wc} and further references therein and also~\cite{Kraus:2013esi,deBoer:2013gz,Compere:2013nba,Ammon:2013hba} for recent progress on holographic thermodynamic and entanglement entropies in this backgrounds) that also admit an extremal configuration of charges.
The second is the violation of the ER=EPR relation~\cite{Maldacena:2013xja} for extremal black holes.
The violation is intuitive from the perspective of the Penrose diagram that is cut in half in the extremal limit, but a thorough analysis of this mechanism in both the gravity and gauge theory setups could shed light on this fascinating proposal.

\subsection*{Acknowledgments}

It is a pleasure to thank Rajesh Gopakumar, Shinji Hirano, Sachin Jain, Gautam Mandal, Ilies Messamah, and Shahin Sheikh-Jabbari for discussions and comments.
This research is supported by the South African Research Chairs Initiative of the Department of Science and Technology and the National Research Foundation. HS is supported by Foundation for Polish Science MPD Programme co-financed by the European Regional Development Fund, agreement no.\ MPD/2009/6.

\appendix

\section{\label{app:A}Non-extremal BTZ} 
For completeness of our presentation, we review the computation of the entanglement entropy in the CFT dual to the non-extremal spinning BTZ~\cite{Hubeny:2007xt,Caputa:2013eka}.
The metric is given by~\eqref{neBTZ}. It can be mapped to the Poincar\'e patch~\eqref{AdS3} via the transformation
\be
w_{\pm} = \sqrt{\frac{r^2-r^2_+}{r^2-r^2_-}}e^{2\pi T_\pm u_\pm} ~, \qquad
y=\sqrt{\frac{r^2-r^2_+}{r^2-r^2_-}}e^{\pi T_+ u_++\pi T_-u_-}\label{mapNE} ~,
\ee
where 
\be
T_{\pm}\equiv\frac{1}{\beta_{\pm}}=\frac{r_+\mp r_-}{2\pi R} ~, \qquad u_{\pm}=\phi\pm \frac{t}{R} ~.
\ee
We make the following observation.
In the BTZ metric, $\phi$ is a periodic coordinate, $\phi\sim\phi+2\pi$, whereas in AdS$_3$, $\phi\in \mathbb{R}$.
In other words, BTZ can be obtained from AdS$_3$ through periodic identifications (see, \textit{e.g.},~\cite{KeskiVakkuri:1998nw}).

Following,~\cite{KeskiVakkuri:1998nw}, we can take the $r\gg r_{\pm}$ limit of \eqref{mapNE} and read off the conformal map between the two two-dimensional CFTs dual to AdS$_3$ and the non-extremal BTZ.
It is given by
\be
w_{\pm}(u_{\pm})=e^{\frac{2\pi}{\beta_{\pm}}  u_{\pm}} ~.
\ee
(For convenience, we set $R=1$.)
This is the map between the plane and a twisted cylinder (for $\beta_+=\beta_-$, it is the usual map from the cylinder to the plane).
Now, we may analytically continue to the imaginary time with periods $\beta_\pm$ and introduce complex coordinates.
This gives us the CFT at finite temperature(s) and finite size.
We will be only interested in the infinite size limit so we consider $\phi\in \mathbb{R}$.

In order to compute the entanglement entropy for a single interval $L$ in the CFT dual to non-extremal spinning BTZ, we use the replica trick and evaluate it through the fixed-time two-point function of the twist fields inserted in the boundary of the entangling interval.
The final formula reads
\be
S_\mathrm{EE}=-\lim_{n\to1}\partial_n\langle \phi^+(z_1,\bar{z}_1)\phi^-(z_2,\bar{z}_2)\rangle ~.
\ee

The two-point correlator on the twisted cylinder can be obtained by inserting \eqref{mapNE} to the general transformation of the CFT correlators \eqref{corr-trans}
and using the appropriately regularized two-point correlator on the plane
\be
\langle {\cal O}(w_1,\bar{w}_1){\cal O}(w_2,\bar{w}_2)\rangle=
\left(\frac{w_2-w_1}{\epsilon}\right)^{-2h}\left(\frac{\bar{w}_2-\bar{w}_1}{\epsilon}\right)^{-2\bar{h}}
\label{plane} ~.
\ee
The conformal dimensions of the twist fields are
\be
h=\frac{c}{24}(n-\frac{1}{n}) ~,\qquad \bar{h}=\frac{\bar{c}}{24}(n-\frac{1}{n}) ~.
\ee
Setting
\be
z_1=\bar{z}_1=0,\qquad z_2=\bar{z}_2=L ~,
\ee
we have the correlator
\bea
\langle \phi^+(0)\phi^-(L)\rangle&=&\left[\frac{\beta_+}{2\pi\epsilon}e^{-\frac{\pi L}{\beta_+}}\left(e^{\frac{2\pi L}{\beta_+}}-1\right)\right]^{-2h} \left[\frac{\beta_-}{2\pi\epsilon}e^{-\frac{\pi L}{\beta_-}}\left(e^{\frac{2\pi L}{\beta_-}}-1\right)\right]^{-2\bar{h}}\nn\\
&=&\left(\frac{\beta_+}{\pi\epsilon}\sinh\frac{\pi L}{\beta_+}\right)^{-2h}\left(\frac{\beta_-}{\pi\epsilon}\sinh\frac{\pi L}{\beta_-}\right)^{-2\bar{h}} ~.
\eea
Differentiating with respect to $n$ and taking the limit $n\to 1$ gives the entanglement entropy
\be
S_\mathrm{EE}=\frac{c}{6}\log\left(\frac{\beta_+}{\pi\epsilon}\sinh\frac{\pi L}{\beta_+}\right)+\frac{\bar{c}}{6}\log\left(\frac{\beta_-}{\pi\epsilon}\sinh\frac{\pi L}{\beta_-}\right) ~.
\ee
The large $L$ (or small $\beta_\pm$) expansion yields 
\be
S_\mathrm{EE}\sim \frac{c\pi L}{6\beta_+}+\frac{\bar{c}\pi L}{6\beta_-}+\frac{c}{6}\log\left(\frac{\beta_+}{2\pi\epsilon}\right)+\frac{\bar{c}}{6}\log\left(\frac{\beta_-}{2\pi\epsilon}\right) ~.
\ee
The small $L$ (or large $\beta_{\pm}$) expansion yields
\be
S_\mathrm{EE}\sim \frac{c+\bar{c}}{6}\log\frac{L}{\epsilon}+\frac{\pi^2 L^2}{36}\left(\frac{c}{\beta^2_+}+\frac{\bar{c}}{\beta^2_-}\right) ~.
\ee

\section{\label{app:B}Energy-momentum tensor for the near-horizon extremal BTZ} \label{Tvv}
In Gaussian normal coordinates an asymptotically AdS$_3$ metric takes the form,
\be
ds^2=dy^2+g_{ij}dx^i dx^j ~.\label{G-met}
\ee
For such a metric it was shown in~\cite{balak, Kraus:2006wn} that the boundary energy-momentum tensor is given by
\be
T_{ij}=-\frac{1}{8\pi G_N}\,\left(K_{ij}-\Tr\, K g_{ij}\right) ~,
\ee
where
\be
K_{ij}=\frac{1}{2}\,\rd_y g_{ij}\,,~~~~\Tr\, K=K_{ij}\,g^{ji} ~.
\ee

The CFT dual to the near-horizon geometry \eqref{NH1} has just a chiral part of a full CFT$_2$ based on $\rd_{v_-}$.
The energy associated to the boundary theory is given by $T_{v_- v_-}$ component of the energy-momentum tensor.
It is often convenient to use another coordinate for this direction with period $2 \pi r_0$.
For this propose we introduce $\varphi=\frac{R}{r_0}\,v_-$

The near-horizon metric \eqref{NH1} has the form~\eqref{G-met}.
We calculate
\begin{eqnarray}
 K_{ij}=\frac{r_0\,e^{\frac{2 y}{R}}}{2}\,\left(\begin{array}{cc}
 0&1\\
 1&0
 \end{array}\right) ~, \qquad \Tr\, K=\frac{2}{R} ~.
 \end{eqnarray}
This gives the stress-energy tensor
\bea
 T_{ij}=\frac{1}{16\pi G_N}\,\left(\begin{array}{cc}
 0&r_0\,e^{\frac{2\rho}{R}}\\
 r_0\,e^{\frac{2\rho}{R}}&4 \frac{r_0^2}{R}
 \end{array}\right) ~.
\eea
Therefore,
\be
T_{\varphi \varphi}=\frac{r_0^2}{4\pi G_N R}~.
\ee
This is the component of energy-momentum tensor that carries the energy. 
This is because $\varphi$ (or $v_-$) is the periodic direction and corresponding excitations are allowed.
It is in agreement with DLCQ approach advocated by~\cite{Balasubramanian:2009bg}.


\end{document}